\documentclass[11pt]{article}
\usepackage{epsfig}
\usepackage[usenames]{color}
\usepackage{amsmath}
\usepackage{amsfonts}
\usepackage{amssymb}
\usepackage{mathrsfs}
\newtheorem{proposition}{Proposition}[section]
\newcommand{\bpr}{\begin{proposition}}
\newcommand{\epr}{\end{proposition}}

\newcounter{Roman}
\addtocounter{Roman}{1}

\newcommand{\beq}{\begin{equation}}
\newcommand{\eeq}{\end{equation}}
\newcommand{\bea}{\begin{eqnarray}}
\newcommand{\eea}{\end{eqnarray}}
\newcommand{\bml}{\begin{multline}}
\newcommand{\bal}{\begin{align}}
\newcounter{saveeqn}

\newcommand{\ssc}{\scriptscriptstyle}

\newcommand{\tr}{{\rm tr}}

\newcommand{\bb}[1]{\mathbb{#1}}

\newcommand{\btr}[1]{|\![#1]\!|}
\input epsf
\begin{document}

\begin{center}{\Large\bf Multigravity and the cosmological constant}\\[2cm] 
{E. T. Tomboulis\footnote{\sf e-mail: tomboulis@physics.ucla.edu}
}\\
{\em Mani L. Bhaumik Institute for Theoretical Physics\\
Department of Physics and Astronomy, UCLA, Los Angeles, 
CA 90095-1547} 
\end{center}
\vspace{1cm}

\begin{center}{\Large\bf Abstract}\end{center}  
The possibility of flat metric solutions in the presence of arbitrary, in particular, Planck mass sized cosmological  constants is investigated within multigravity theory, the ghost-free theory of $(N+1)$ rank-2 tensor fields comprising one massless and $N$ massive spin-2 fields. It is found that, indeed,  for any $N$, no matter how large, no flat solutions exist without fine-tuning parameters in the potential. For an infinite number of fields, however, flat solutions exist within an infinite-dimensional parameter space. Such a ground state  may be viewed as a collective state that cannot be approximated by any finite $N$.

\vfill
\pagebreak

The cosmological constant (c.c.) problem is one of the fundamental  unsolved problems of present-day physics (cf. \cite{C} for a review). 
The observed vacuum energy is $\rho_V \sim  (10^{-12}{\rm Gev})^4$.  
If one defines the c.c. $\Lambda$ in the standard manner through the term $ (1/16\pi G) \Lambda \int d^4 x \sqrt{-g}$ 
in the gravity action, the observed $\rho_V$  correspond to a c.c. $\Lambda \sim 10^{-84} {\rm Gev}^2$. Quantum field theory (QFT), on the other hand, naturally yields c.c.  of the order of the cutoff scale below which the theory remains valid. Quantized matter fields in a classical gravitational background thus result in c.c. that are huge compared to the observed value. As General Relativity (GR) appears to be a good effective field theory up to scales of the order of the Planck mass,  
and local QFT is believed to also remain valid approximately up to such scales, one generally expects c.c. of the order of $\Lambda \sim (10^{18} {\rm Gev})^2$. This is the  infamous ``120 orders of magnitude discrepancy problem".
It is a baffling problem as it actually occurs at all scales, i.e., at whatever scale (cutoff) may be taken as characteristic of the matter system. 
Furthermore, any phase transition within a given system, such as the chiral or deconfinement transitions in QCD, or the electroweak transition in the Higgs sector, results in contributions to $\Lambda$ that are independent of any ultimate UV cutoff that may apply to the system.  These contributions are already huge compared to the observed value. 

Since the observed $\Lambda$ is so tiny, it is natural to seek flat metric solutions, which are subsequently subject to some small perturbations. The problem is, of course, that flat solutions generically require extreme fine-tuning of the parameters in the action - this in fact may be taken as the field theoretic  statement of the c.c. problem.  
Here we investigate the possibility of flat metric solutions in multigravity theory (cf. \cite{dR}, \cite{Hint} for reviews). 
The construction of the massive gravity potential \cite{dRGT1}, \cite{HR1}, \cite{HR2}, \cite{dRGT2}, \cite{HRS-M}, \cite{HR3}  solved the problem of a consistent interacting massive spin-2 theory evading the famous BoulwareDeser ghost \cite{BD}. It was subsequently extended first to bigravity \cite{HR4} and then multigravity \cite{HintR}, the general coordinate invariant theory of $N+1$ rank-2 fields comprising one massless and $N$ massive spin-2 fields.  
We find that, indeed, for any finite number of fields $N$, no matter how large, no flat solutions are possible without fine-tuning of action parameters. For an infinite number of fields, however, there are flat solutions. Such ground state solutions may be viewed as a kind of a collective state that cannot be approximated by any finite $N$.

\bigskip

The multigravity action is defined by adding to 
the familiar Einstein - Hilbert terms for each rank-2 tensor field $f_i$, $i=0, \ldots, N$, the potential terms  $U^{(i)}$ 
pairwise coupling adjacent fields $f_i, f_{i+1}$,  $i=0, \ldots, N-1$. They are functions of the  matrix 
\beq
\bb{X}_{i,i+1} \equiv \sqrt{f_i^{-1}f_{i+1}}   \; .    \label{Xmatrix}
\eeq
The square root is understood in the usual matrix sense:  
   $(f_i^{-1}f_{i+1})^\mu_{~\nu}  =    X^\mu_{~\alpha} X^\alpha_{~\nu}$. 
The general form of the potential $U^{(i)}$ (in dimensionless units) is explicitly given by 
\beq 
U^{(i)}(\bb{X}_{i,i+1}) =\sqrt{-f_i} \sum_{l=0}^4 \beta^{(i)}_l {\rm S}_l(\bb{X}_{i,i+1})  \; ,   \label{act1}
\eeq 
where the coefficients $\beta^{(i)}_l$ are arbitrary real numbers.   
The $\rm S_l$ are the symmetric polynomials constructed out of the eigenvalues 
of the matrix $\bb{X}$; they can be expressed in terms of the traces of powers of $\bb{X}$. In $d=4$, with the common notation  $\btr{\bb{X}^k} \equiv \tr \bb{X}^k$, 
they are: 
\beq
{\rm S}_0 = 1, \;  
{\rm S}_1 = \btr{\bb{X}},   \;      
{\rm S}_2  = {1\over 2!} \big( \btr{\bb{X}}^2  -  \btr{\bb{X}^2]}\big) , \;  
{\rm S}_3  = {1\over 3!} \big( \btr{\bb{X}}^3 - 3\btr{\bb{X}}\btr{\bb{X}^2} + 2\btr{\bb{X}^3}    \big),  \;
{\rm S}_4 = \det \bb{X}        \label{act2}
 \, .   
\eeq
In (\ref{act1}) 
the $\beta^{(i)}_0$ term is a cosmological constant term for $f_i$, whereas, by (\ref{act2}), the $\beta^{(i)}_4$ term is a cosmological term for $f_{i+1}$. When  adding the potential terms $U^{(i)}(\bb{X}_{i,i+1})$, $i=0, \ldots, N-1$, however, the $\beta^{(i)}_4$ term in $U^{(i)}(\bb{X}_{i,i+1})$ can be absorbed in the  
$\beta^{(i+1)}_0$ term of $U^{(i+1)}(\bb{X}_{i+1,i+2})$ {\it except} for the last $\beta^{(N-1)}_4$ term. Relabeling
$\beta^{(N-1)}_4 \equiv \beta^{(N)}_0$, the action is then given by 
\beq
S=\sum_{i=0}^{N} 
\frac{1}{\kappa^2}\! \int \!d^dx \! \sqrt{-f_i} R(f_i)   - 
\frac{1}{\kappa^2} \! \int \! d^dx \!  \Bigg\{  \sum_{i=0}^{N-1}  M_i^2 \! \Big( \sqrt{-f_i} \sum_{l=0}^3 \beta^{(i)}_l {\rm S}_l(\bb{X}_{i,i+1})  \Big) +   \sqrt{-f_N} M_{N-1}^2 \beta^{(N)}_0 \Bigg\}    \label{act3}
\eeq
with $\kappa^2= 16\pi G$.\footnote{We could allow different couplings $( \kappa^2/ \gamma_i)$, $\gamma_i >0$, for each $f_i$ but such generalizations make no qualitative difference in the following. \label{f0}}
The masses  $M_i$ are taken to be of the order of Planck mass or higher. Having set the scale of each  potential term by $M_i$, all
$\beta^{(i)}_l$ are taken to be comparable in magnitude and of order  unity. It would indeed be unnatural to take  $\beta^{(i)}_l$ for different $l$ varying  greatly in magnitude.

It is clear from (\ref{act3}) that a flat metric solution $\bar{f}_i = \eta$, all $i$, is not possible without 
fine-tuning the parameters $\beta_l^{(i)}$ by imposing relations among them.  Thus, in $d=4$, one must set $\beta^{(0)}_0=( -3 \beta^{(0)}_1 - 3 \beta^{(0)}_2 -\beta^{(0)}_3)$ and also fix the other 
$\beta^{(i)}_0$, $i=1, \ldots, N$, by similar linear relations.

\bigskip

We now seek solutions of the form\footnote{Note that because of the ratio form of $\bb{X}_{i, i+1}$,  one may just  set $C_0=1$. \label{f1}} 
\beq
\bar{f}_i = \overline{C}_i^2  \bar{g} \,, \quad  \mbox{with} \quad \overline{C}^2_i =\prod_{j=0}^i C_j^2 \, , \quad C_0=1\,, 
\qquad i=0, \ldots, N \; , 
\label{soln1}
\eeq
for some metric $\bar{g}$ and real constants $C_i$. 
Due to the unwieldy square root form of (\ref{Xmatrix}) the equations of motion (e.o.m.) following from (\ref{act3}), though straightforward to derive, are somewhat involved. For fields of the assumed form (\ref{soln1}), however, one simply has $\bb{X}_{i,i+1}= {\bf{1}} \bar{C}_i^{-1} \bar{C}_{i+1} = {\bf 1} C_{i+1}$ for any $\bar{g}$. 
Taking for the moment $\bar{g}$ to be flat, $\bar{g} = \eta$, the e.o.m. in $d=4$ become the following algebraic equation system for the $C_i$: 
\bal 
&\beta^{(0)}_0 + 3\beta^{(0)}_1C_1 + 3 \beta^{(0)}_2 C^2_1 + \beta^{(0)}_3 C^3_1 =0 \; . \label{eom1}\\ 
& \frac{M^2_{i-1}}{M^2_i} \Bigg( \beta^{(i-1)}_1 C_i^{-3} +  3 \beta^{(i-1)}_2C_i^{-2} +  3\beta^{(i-1)}_3 C_i^{-1} 
\Bigg)  \nonumber  \\  
& \qquad \qquad + \;   \beta^{(i)}_0    +  3\beta^{(i)}_1 C_{i+1} +  3\beta^{(i)}_2 C_{i+1}^2  +\beta^{(i)}_3 C_{i+1}^3  = 0 
\, , \qquad \quad   i=1, \dots, (N-1)  \;. \label{eom2} \\
&  \mbox{and}              \nonumber \\
& \beta^{(N-1)}_1 C_N^{-3} +  3 \beta^{(N-1)}_2 C_N^{-2} +  3\beta^{(N-1)}_3 C_N^{-1}  + \beta^{(N)}_0  =0 \; . \label{eom3} 
\end{align} 
(\ref{eom1}) - (\ref{eom3}) is an algebraic  system of cubic equations, which is already decoupled. 
(\ref{eom1}), from the e.o.m. for $f_0$, determines $C_1$; (\ref{eom2}), from the e.o.m. for $f_i$, $i=1, \ldots, N-1$,  determine then successively $C_2, C_3, \cdots, C_N$. In fact they can each be solved analytically by the 
Cardano-Tartaglia (GT) formula for the solution of the cubic equation.\footnote{One may, of course,  consider less general models for illustrative purposes, e.g., the `minimal model' of \cite{HR4}, defined by setting $\beta^{(i)}_2 = \beta^{(i)}_3 = 0$. In this case (\ref{eom1})-(\ref{eom3}) becomes a linear system.\label{f2} } Note that with the coefficients $\beta^{(i)}_l$ of `order unity', the $C_i$ solving the system (\ref{eom1})-(\ref{eom2}) are generally also of that order and, by this iterative procedure, each $C_{i+i}$ depends on the set $\{ \beta^{(k)}_l\}$,
 $k\leq  i$. Furthermore, for generic parameters 
$\{\beta^{(j}_l\}$, the solutions $C_i \neq  0$, since $C_i=0$ requires the vanishing of a discriminant, which requires relations among $\beta^{(j)}_l$'s.  It is important that, by the GT formula, there is always at least one real solution. 

This leaves (\ref{eom3}) from the e.o.m. for $f_N$. Since all $C_i$ have been determined by (\ref{eom1}) - (\ref{eom2}), it is clear that this equation cannot be satisfied unless $\beta^{(N)}_0$ is fine-tuned. 
Alternatively, one may take  $\bar{g}$ to be a metric of constant curvature $\lambda$. In the latter case, the system (\ref{eom1}) - (\ref{eom3}), now augmented by the additional terms from the variation of the EH terms in (\ref{act3}), determines the $(N+1)$ parameters $C_i, \lambda$. With mass parameters of Planck order, however, the residual c.c. is of the same magnitude, so nothing is gained. 

Returning to the $\bar{g} = \eta$ case, one may note that, actually, something has been gained by using the Ansatz 
(\ref{soln1}) versus simply setting $\overline{f}_i =\eta$ for all $i$: instead of having to fine-tune all $\beta^{(i)}_0$,  
only $\beta^{(N)}_0$ has to be fine-tuned. All other $\beta^{(i)}_l$ remain arbitrary while all the e.o.m. are satisfied. This one fine-tuning may be avoided by augmenting the last term in (\ref{act3}) to a full potential term $U^{(N)}(\bb{X}_{N,N+1})$ by  the addition of a field $f_{N+1}$: the e..o.m. for $f_N$ now determines $C_{N+1}$. 
This of course transfers the problem to fine-tuning the new $\beta^{(N+1)}_0$ parameter, which can be evaded by adding the full $U^{(N+1)}(\bb{X}_{(N+1,N+2)})$ term, and so on ad-infinitum. A formal flat space solution is thus obtained by replacing (\ref{act3}) by: 
\beq
S=\sum_{i=0}^{\infty} {1 \over \kappa^2}\int d^dx  \sqrt{-f_i}\, R(f_i)   - 
\frac{1}{\kappa^2} \Bigg\{  \sum_{i=0}^{\infty}  M_i^2 \! \int d^dx \, \Big( \sqrt{-f_i} \sum_{l=0}^3 \beta^{(i)}_l {\rm S}_l(\bb{X}_{i,i+1})  \Big) \Bigg\}  \; .    \label{act4}
\eeq
It is crucial that an infinite tower of $f_i$ fields is present. Any truncation to finite tower leaves a large residual c.c, i.e., a discrepancy away from the flat solution.

There are other examples where such a situation arises. An elementary, but fundamental, example is provided by the canonical commutation relation of non-relativistic QM: ${\bf XP - PX} = i\hbar {\bf 1}$. 
This can hold only for infinite matrices representing ${\bf X, P}$, since, as obvious at a glance, this relation is mathematically nonsensical for any finite dimensional matrices. Given  matrices for ${\bf X, P}$ obtained from the e.o.m. for the system, restricting to any $N\times N$ submatrix necessarily creates a discrepancy between the two sides that grows with $N$.\footnote{This is well illustrated by the exact 
harmonic oscillator solution in the energy representation  (the original matrix QM).  
Note that in  the coordinate representation,  where matrices have continuously labeled indices, $< x^{\prime\prime} | {\bf X} | x^\prime> =x^\prime \delta(x^{\prime\prime} - x^\prime)$, the question of a finite dimensional approximation does not arise. \label{f3}} To avoid the discrepancy one has to enlarge to $(N+1)\times (N+ 1)$, which in turn necessitates $(N+2)\times (N+ 2)$  matrices and so on ad infinitum. This is then closely analogous to the case here, where the flat solution is obtained only by extending to an infinite tower of $f_i$ fields. 

\bigskip

The infinite tower of fields $f_i$ then collectively adjust their individual ground states $\bar{f}_i$, which are correlated by the e.o.m., 
so that any c.c. in (\ref{act4}) are absorbed leaving a flat metric. 
To better understand this solution consider expanding about it: 
$f_i = \overline{f}_i + \chi_i  = \overline{C}_i^2 \eta + \chi_i $,  $i=1,2, \ldots \ $.   
Substituting in (\ref{act4}), upon expanding  
the terms linear in $\chi_i$ vanish since $\bar{f}_i$ satisfy the e.o.m. 
Noting that $\sqrt{-f_i} = \overline{C}_i^4 \det [ \bf{1} +  \overline{C}_i^{-2} \eta^{-1} \chi_i ]$,  
the part of the EH action quadratic in $\chi_i$ is of the form: 
\beq
S_{\ssc (2)}^{\ssc\rm EH} =\sum_{i=0}^\infty \frac{1}{\kappa^2}  \overline{C}^{-2}_i \int d^4x \; \partial_{\alpha} \chi_{i \kappa\lambda} K^{\alpha\kappa\lambda\beta\mu\nu}(\eta^{-1})\partial_\beta \chi_{i\mu\nu} \; . \label{EHact1}
\eeq
In (\ref{EHact1}) $K^{\alpha\beta\kappa\lambda\mu\nu}(\eta)$ is the usual EH kinetic tensor on flat metric. 
To obtain canonically normalized quadratic kinetic terms then, each $\chi_i$ has to be rescaled: $\chi_i \to \kappa \overline{C}_i$, so that the expansion assumes the form: 
\beq
f_i =  \overline{C}_i^2 \eta + \kappa \overline{C}_i\chi_i =  \overline{C}_i^2 \eta \, [ 1 + \kappa \overline{C}_i^{-1} \eta^{-1} \chi_i ] 
\;, \qquad i=1,2, \ldots   \; .   \label{exp1}
\eeq
The interactions are obtained by 
expansion of the ``$\Gamma\Gamma$-form" of the EH action, which gives the most efficient form for obtaining the interaction vertices (for a recent discussion cf. \cite{TT1}). It 
is obtained by simply multiplying by $\ \det [ \bf{1} +  \overline{C}_i^{-1} \eta^{-1} \chi_i ]$ and replacing $K(\eta^{-1})$ by $K([ 1 + \kappa \overline{C}_i^{-1} \eta^{-1} \chi_i ]^{-1} )$ in the rescaled (\ref{EHact1}). The EH interactions are then: 
\beq 
S_{\ssc \rm I}^{\ssc\rm EH} = \sum_{i=0}^\infty  \, \sum_{n\geq 3}^\infty  \kappa^{n-2} \int d^4x \, V^{\alpha \beta \mu_1\nu_1\mu_2\nu_2 \cdots \mu_n \nu_n}_{i,n}( \eta^{-1})  \partial_\alpha \chi_{i\mu_1\nu_1} \partial_\beta \chi_{i\mu_2\nu_2} 
\overline{C}^{-1}_i \chi_{i\mu_3\nu_3} 
\ldots \overline{C}^{-1}_i \chi_{i\mu_n\nu_n} \; ,  \label{EHact2} 
\eeq 
where the $V_{i,n}$ are the usual EH $n$-point interaction vertices for the field $\chi_i$. 

The expansion is then an expansion in $\kappa \overline{C}_i^{-1}$. The products $\overline{C_i}$, (\ref{soln1}), are fully determined as functions of $\{\beta^{(i)}\}$ by the e.o.m. as detailed above. There are three possible behaviors for $\overline{C}_i$  as $i \to \infty$:  (i) $\overline{C_i} \to \infty$; (ii) $\overline{C_i} \to 0$; or  (iii) $\overline{C_i} \to {\rm constant}$. These correspond to  the solution $C_i$ to (\ref{eom2}) being: (i) $C_i > 1$; (ii) $C_i < 1$ and (iii) $C_i =1$, respectively, for each $i\geq N_0$ for some finite $N_0$, which may be arbitrarily large.  
(iii) implies relations between the potential parameters $\beta^{(i)}_k$ for each $i\geq N_0$ and thus fine-tuning. As seen from the CT  formula, (i) and (ii) can be satisfied by an infinite subset of $\{ \beta^{(i)}_l\}$ of the same cardinality (order of infinity) as the set $\{ \beta^{(i)}_l\}$. (ii) has to be rejected since it would lead to arbitrarily low masses and hence not viable range of validity as an effective theory.  Thus, (i) is the only possible option. It implies an infinite tower of spin-2 fields with rising mass spectrum (see below).  

Noting that $\bb{X}_{i,i+1}= C_{i+1}\Big[ [{ \bf 1} +\kappa \overline{C}_i^{-1} \eta^{-1} \chi_i ]^{-1}  [ {\bf1} + \kappa \overline{C}_{i+1}^{-1} \eta^{-1} \chi_{i+1} ] \Big]^{1/2}$, 
the expansion of the potential terms gives the quadratic mass term
\beq
S^{\ssc \rm P}_{(2)} = M_0^2  \int d^4x\,   \mathcal{M}^{\kappa\lambda\mu\nu}_{\ssc \rm FP}(\eta^{-1}) \,  \sum_{i,j=0}^\infty \chi_{j \kappa\lambda} \bb{M}_{ji} 
\chi_{i\mu\nu} \; , \label{Potact1}
\eeq
where $\mathcal{M}^{\kappa\lambda\mu\nu}_{\ssc FP}(\eta^{-1}) =\Big(\eta^{\kappa\mu} \eta^{\lambda\nu} -\eta^{\kappa\lambda}  \eta^{\mu\nu}\Big)$ is the Fierz - Pauli tensor structure. 
The (dimesionless) mass matrix $\bb{M}_{ji}$ is a real symmetric tri-diagonal infinite matrix.  
Its elements have the form  $\bb{M}_{ji} = \overline{C}_j b_{ji} \overline{C}_i$ with $b_{ji}=b_{ij}$ and $b_{ji}=0$ whenever $|i-j| \geq 2$.  
 The $b_{ij}$ are functions of the parameters $a_k \equiv (M^2_k/M^2_0)$ and the set of $\beta^{(k)}_l$ coefficients with $k \leq i$.\footnote{The quantities $b_{ji}$, as obtained via the expansion (\ref{exp1}), are expressed in terms of  $\{\beta^{(k)}\}$ and $\{C_{k+1}\}$, but, of course, the latter are included as $\beta^{(k)}$ dependence since all $C_{k+1}$  have been fixed by the e.o.m (\ref{eom1}) - (\ref{eom2}) as functions of the $\{\beta^{(k)}\}$,  $k\leq i$. \label{f4} } 
 
Expansion of the potential beyond quadratic order gives the potential interactions terms of the general form:  
\bml
S^{\ssc \rm P}_{\ssc \rm I} = M_0^2 \sum_{i=0}^\infty \sum_{n\geq 3} \kappa^{n-2} \int d^4x \;  \mathcal{V}^{\mu_1\nu_1\mu_2\nu_2\cdots \mu_n\nu_n}_n\Big(\eta^{-1}; \{a_k, \beta^{(k)}\}_{k\leq i}\Big) \overline{C}_{i_1} \chi_{i_1 \mu_1\nu_1} 
 \overline{C}_{i_2} \chi_{i_2 \mu_2\nu_2} \\ 
 \qquad \quad \cdot \overline{C}^{-1}_{i_3} \chi_{i_3 \mu_1\nu_3} \cdots 
 \overline{C}^{-1}_{i_n} \chi_{i_n \mu_1\nu_n} \, ,  \;  \qquad  i_j \in \{i, i+1\} \; . \label{Potact2}
\end{multline}

For a finite system of $N+1$ fields $\chi_i$ the  tridiagonal mass matrix $\bb{M}_{ji}$ is a non-negative definite matrix possessing  one zero and $N$ non-zero eigenvalues. This follows from the general Hamiltonian analysis in \cite{HR3}, \cite{HR4}, \cite{HintR} and we assume that it holds in the limit $N\to \infty$. The mass matrix $\bb{M}_{ji}$ is  real symmetric and hence, trivially, satisfies $\bb{M}_{i,i+1} \bb{M}_{i+1,i} > 0$, which, for a tridiagonal matrix, implies that the eigenvalues are distinct \cite{MM}.  This property then  implies a rising eigenvalue spectrum. 
 Note that this means that one has a tower of rising mass eigenvalues even if one sets $a_i =1$, i.e., $M_i^2 = M_0^2$, all $i$. 
This in fact makes little difference as the rise is dominated by the $\overline{C}_i$ factors in the $\bb{M}$ matrix elements, which grow asymptotically very rapidly. The size of the eigenvalues can be estimated by taking finite `sections' (submatrices)  \cite{L}, \cite{SSZ} in conjunction with the theory of Ger\v sgorin disks \cite{MM}. 
Listing the eigenvalues $\lambda^{(k)}$ in increasing order, one has the estimate $\lambda^{(k)} \lesssim |\overline{C}_k \overline{C}_{k+1}|$.

The mass matrix $\bb{M}$ is diagonalizable by a real  orthogonal matrix $U$, whose $k$-th column is the column eigenvector $e^{(k)}$ belonging to eigenvalue $\lambda^{(k)}$. Then, 
with $\chi$ and $\phi$ denoting the column vectors of the $\chi_i$ fields and the mass eigenstate fields $\phi_i$, $i=0,1,2,\ldots$, respectively, one has $\chi_i = \sum_{j=0}^\infty e^{(j)}_i\phi_j$ and $\phi_i= \sum_{j=0}^\infty e^{(i)}_j\chi_j$. Inserting unity in the form $(U^{T}U)_{ij} = \delta_{ij}$ in the canonically normalized EH kinetic term, i.e., in (\ref{EHact1}) after the rescaling $\chi_i\to \kappa \overline{C}_i \chi_i$, one obtains the quadratic part of the action 
in terms of the mass eigenstate fields:
\beq
 S^{\ssc \rm EH}_{(2)} + S^{\ssc \rm P}_{(2)} = \sum_{i=0}^\infty  \int d^4x \; \left[ \partial_{\alpha} \phi_{i \kappa\lambda} K^{\alpha\kappa\lambda\beta\mu\nu}(\eta^{-1})\partial_\beta \phi_{i\mu\nu}   
 +    \mathcal{M}^{\kappa\lambda\mu\nu}_{\ssc \rm FP}(\eta^{-1}) \, \mu^2_i  \phi_{i \kappa\lambda} \phi_{i\mu\nu}  \right]\; 
  \label{quadract}
\eeq
with $\mu^2_i = M_0^2 \lambda^{(i)}$. The interaction parts  can then be expressed in terms of the mass eigenstate fields by substituting $\chi_i = \sum_{j=0}^\infty e^{(j)}_i\phi_j$ in (\ref{EHact2}), (\ref{Potact2}).  

\bigskip

In summary, we have  obtained a flat solution of the form (\ref{soln1}) in multi-gravity with an infinite number of fields. As we saw it is absolute crucial that the number of fields be infinite in order to obtain a flat solution, i.e., $\bar{g} = \eta$ in (\ref{soln1}).  The  solution holds for arbitrary, in particular, Planck-mass-sized cosmological constants for each field $f_i$, the scale being set by the parameters $M^2_i$ in the action (\ref{act4}), and in the infinite-dimensional space of the parameters $\{\beta^{(i)}_l\}$. 
Such cosmological constants are expected to result from integrating out matter fields in the multigravity  background and are indeed necessary in order that multi-gravity holds as an effective field theory up to Planck order scales (just as GR does).   

The flat solution (\ref{soln1}): $\bar{f}_i = \overline{C}_i^2  \eta$, $i=0,1,2,\ldots$, (\ref{soln1}), is the ground state of the infinite collection of fields $f_i$. The constants $\{C_i\}$ are determined by the e.o.m (\ref{eom1}), (\ref{eom2}) in terms of the parameters in the action. This ground state may be viewed as a coherent state that, as we saw, cannot be approximated by any finite number of fields, no matter how large.  
It collectively adjusts to accommodate any changes in the action parameters maintaining a flat solution. Expanding about this ground state gives a well-ordered perturbative expansion. Expressed in terms of mass eigenstates upon diagonalization of the mass matrix, it gives an interaction theory   of  a massless spin-2  and a tower of massive spin-2 excitations with  asymptotically rapidly rising mass spectrum. Since all masses are around  Planck mass or higher, this implies that, for all practical purposes, almost all these excitations above this ground state decouple  and one is left with a massless spin-2 and one or two massive spin-2 around Planck scale. 

Were one to be given this end result  without knowing where it came from, one would naturally view it as the usual   manifestation of the cosmological constant problem: a number of excitations propagating on a flat background obtained from a potential dependent on seemingly fine-tuned parameters $C_i$  and $\hat{\beta}^{(i)}_l$. But these parameters are not fine-tuned: $\hat{\beta}^{(i)}_l = \beta^{(i)}_l C_{i+1}^l$ and $\{C_i \}$ are determined by the e.o.m (\ref{eom1}), (\ref{eom2}) for arbitrary  parameters $\beta^{(i)}_l$ in the starting action (\ref{act4}) as detailed above. 

An infinite number of fields would indicate extended structure/nonlocality in whatever fundamental theory underlies (\ref{act4}) viewed as an effective theory. 
There have been some attempts to extract massive gravity or bigravity from string theory, e.g \cite{LMMS}, but so far they remain inconclusive.

The observed c.c. is of course not zero but extremely tiny. It is a striking fact that it corresponds to cosmological mass (length) scales of the order of the Hubble constant $\sim (10^{-42} {\rm Gev})$. This suggests that it might arise due to some long distance  mechanism  operating on top of the flat space solution, the latter absorbing all  c.c. contributions due to quantum fluctuations at short (molecular and shorter $\gtrsim 10^{-3} {\rm ev})$ scales. What such a  mechanism might be, though, cannot be discussed here.

\end{document}